\makeatletter
\@ifundefined{@parse@version@dash}{%
\def\@parse@version#1{\@parse@version@0#1}
\def\@parse@version@#1/#2/#3#4#5\@nil{%
\@parse@version@dash#1-#2-#3#4\@nil}
\def\@parse@version@dash#1-#2-#3#4#5\@nil{%
  \if\relax#2\relax\else#1\fi#2#3#4 }
}{}
\makeatother

\documentclass[%
 preprint,
superscriptaddress,
 amsmath,amssymb,
 aps,
]{revtex4-2}

\usepackage{graphicx}% Include figure files
\usepackage{dcolumn}% Align table columns on decimal point
\usepackage{bm}% bold math
\usepackage{relsize}
\usepackage{natbib}
\setcitestyle{authoryear,open={(},close={)}}

\usepackage{xspace} 
\draft % marks overfull lines with a black rule on the right
\usepackage[colorlinks=true,linkcolor=blue]{hyperref}
\usepackage{float}
\usepackage{color}
\usepackage[dvipsnames]{xcolor}
\usepackage{soul}
\usepackage[normalem]{ulem} % allow the strikethrough command \sout
\usepackage[toc,page]{appendix}

\begin{document}

%\reprint{APS/123-QED}

\title{THz emission from Fe/Pt spintronic emitters with L1$_{0}$-FePt alloyed interface}



%\thanks{}

\author{Laura Scheuer}
\author{Moritz Ruhwedel}
\affiliation{Fachbereich Physik and Landesforschungszentrum OPTIMAS, Technische Universit\"{a}t Kaiserslautern, Kaiserslautern 67663, Germany}
\author{Dimitris Karfaridis}
\author{Isaak G. Vasileiadis}
\affiliation{Department of Physics, Aristotle University of Thessaloniki, Thessaloniki 54124, Greece}
\author{Dominik Sokoluk}
\affiliation{Fachbereich Elektro-Informationstechnik and
Landesforschungszentrum OPTIMAS, Technische Universit\"{a}t Kaiserslautern, Kaiserslautern 67663, Germany}
\author{Garik Torosyan}
\affiliation{Photonic Center Kaiserslautern, Kaiserslautern 67663, Germany}
\author{George Vourlias}
\author{George P. Dimitrakopoulos}
\affiliation{Department of Physics, Aristotle University of Thessaloniki, Thessaloniki 54124, Greece}

\author{Marco Rahm}
\affiliation{Fachbereich Elektro-Informationstechnik and
Landesforschungszentrum OPTIMAS, Technische Universit\"{a}t Kaiserslautern, Kaiserslautern 67663, Germany}

\author{Burkard Hillebrands}
\affiliation{Fachbereich Physik and Landesforschungszentrum OPTIMAS, Technische Universit\"{a}t Kaiserslautern, Kaiserslautern 67663, Germany}

\author{Thomas Kehagias}
\affiliation{Department of Physics, Aristotle University of Thessaloniki, Thessaloniki 54124, Greece}

\author{Ren\'{e}  Beigang}
\affiliation{Fachbereich Physik, Technische Universit\"{a}t Kaiserslautern, Kaiserslautern 67663, Germany}

\author{Evangelos Th. Papaioannou}
\email[]{Lead Contact: evangelos.papaioannou@physik.uni-halle.de}
\affiliation{Institut f\"{u}r Physik, Martin-Luther Universit\"{a}t Halle Wittenberg, Von-Danckelmann-Platz 3, 06120 Halle, Germany}

%\authorinfo{Correspondence to Evangelos Papaioannou, email: evangelos.papaioannou@physik.uni-halle.de}
%\thanks{Correspondence to Evangelos Papaioannou, E-mail: evangelos.papaioannou@physik.uni-halle.de}


%\date{\today}

\keywords{THz optics, ultrafast spin dynamics, spin mixing conductance, THz spintronic emitters,L1$_{0}$-FePt}

\begin{abstract}

Recent developments in nanomagnetism and spintronics have enabled the use of ultrafast spin physics for terahertz (THz) emission. Spintronic THz emitters, consisting of ferromagnetic FM /
non-magnetic (NM) thin film heterostructures, have demonstrated impressive properties for the use in THz spectroscopy and have great potential in scientific and industrial applications. In this work, we focus on the impact of the FM/NM interface on the THz emission by investigating Fe/Pt bilayers with engineered  interfaces. In particular, we intentionally modify the Fe/Pt interface by inserting an ordered L1$_{0}$-FePt alloy interlayer. Subsequently, we establish that a Fe/L1$_{0}$-FePt (2\,nm)/Pt configuration is significantly superior to a Fe/Pt bilayer structure, regarding THz emission amplitude. The latter depends on the extent of alloying on either side of the interface. The unique trilayer structure opens  new perspectives in terms of material choices for the next generation of spintronic THz emitters.

\end{abstract}

%\pacs{}% insert suggested PACS numbers in braces on next line

\maketitle

\section*{Introduction}
%\label{}

Ultrafast spin-to-charge conversion in heterostructures composed of ferromagnetic (FM)/non-magnetic (NM) thin films can give rise to the emission of THz electromagnetic waves~\citep{Seifert2016,ADOM:ADOM201600270}. The experimental scheme involves the use of femtosecond (fs) laser pulses to trigger ultrafast spin and charge dynamics in FM/NM bilayers, where the NM layer features a strong spin-orbit coupling~\citep{Kampfrath2013,Papaioannou2021}. Via the inverse spin Hall effect (ISHE) the spin current generated in the FM layer by the fs-laser pulse, is converted  to an ultrashort charge current burst that gives rise to the THz radiation. The THz emission from these spintronic THz emitters (STE) have remarkable properties in terms of signal strength and bandwidth, they are easy to use, robust and do not require electrical connections. Their potential for technological applications is large, while the rich physics behind the excitation and emission has attracted scientific attention~\citep{Singh2021,Jacques2021,Zhang2020,Wolfgang2021}. During the last years, THz spintronic emitters have been heavily investigated aiming to obtain large signals and spectral bandwidths and to incorporate them into THz applications~\citep{Papaioannou2021,Thomas2021,Benjamin2021}. 
In particular, a wide range of material properties have been studied including: different material compositions of FM/NM layers with a variety of  thicknesses~\citep{Torosyan2018,Seifert2016,ADMA:ADMA201603031,ADOM:ADOM201600270,Qiu:18}, ferrimagnetic/NM structures~\citep{spin2017,Albrecht2018,Albrecht2019,Albrecht2020}, and antiferromagnetic metal/NM \citep{Ogasawara_2020}. In addition, the impact of material interfaces were studied by inserting non-magnetic interlayer material such as Cu,Al, Ti, Au and ZnO layers in FM/X/NM trilayers~\citep{Oliver2021,Papaioannou2018,Seifert_2018,Jacques2021,Li_2018}. 
Furthermore, the role of interface engineering at the FM/NM interface was examined~\citep{Sasaki,Li_2019,Papa2019}. It was revealed in Fe/Pt bilayers~\citep{Papa2019} that the performance of STEs can be controlled by structurally optimizing the FM/NM interface quality and its defect density: different defect densities result in changing the elastic electron-defect scattering lifetime in the FM  and NM layers and the interface transmission for spin-polarized, non-equilibrium electrons. A decreased defect density increases the electron-defect scattering lifetime and results in a significant enhancement of the THz-signal amplitude and modifies the spectrum. Further optimization of STE was achieved by alternating  the stack geometry of the STE: cascading FM/NM layers as multilayers~\citep{ADMA:ADMA201603031}, using stacking sequence as NM1/FM/NM2 where NM1 and NM2 have opposite spin Hall angles~\citep{Seifert2016}, inserting the STE layer as an interlayer into a metal-dielectric photonic crystal~\citep{Haifeng2018}, and using metallic trilayer structures with different patterned structures, interface materials and substrates~\citep{doi:10.1002/pssr.201900057,Li_2018,Seifert_2018,Li_2019,Garik2020,Hibberd2019,Kong2019,Spie2020}. Research effort has also been focused on a better utilization of the laser pulse energy by using different excitation wavelengths~\citep{Papaioannou2018,Herapath2019} and exploring the optical damage threshold~\citep{Kumar2021}.

This work focuses on the material aspect of the STE. In contrast to many other research efforts, that have addressed the choice of the materials, we introduce a different direction: the concept of inducing controlled alloyed interlayers at the FM/NM interface. We use Fe/Pt bilayers and modify the interface by controlling the growth temperature of the Pt layer. At specific growth temperatures, an ordered L1$_{0}$-FePt interlayer appears. The Fe/L1$_{0}$-FePt/Pt trilayer amplifies the THz emission by almost a factor of two compared to Fe/Pt bilayers. 

\section*{Results}

\subsection*{Growth }

Fe/Pt bilayers with nominal thicknesses Fe (12\,nm) / Pt (6\,nm) were grown on MgO (100) substrates by electron-beam evaporation in an ultrahigh vacuum (UHV) chamber with a base pressure of 5 $\times$ 10$^{-9}$\,mbar. Aiming at a constant initial total thickness over all the sample series, a modification of the Fe/Pt interface quality was induced by variation of the growth temperature of the Pt layer.\\
The cleaning protocol of the MgO (001) 10 $\times$ 10 mm$^{2}$ substrates involved ex-situ chemical cleaning with acetone and isopropanol and in-situ annealing at 600\,$^\circ$C. 
The incident molecular beam was aligned perpendicular to the MgO substrate with a growth rate in the range of $R_{Fe,Pt}$ = 0.005\,nm/s, controlled by a quartz crystal oscillator during the deposition procedure. The 12\,nm thick Fe film was deposited onto the MgO substrate which was heated at 300\,$^\circ$C  and subsequently annealed at the same temperature for 30\,min. This growth process of Fe has been proven to provide the highest quality of Fe films grown in our UHV system~\citep{Keller2018}. 

On top of the fully formed and annealed Fe film, 6\,nm of Pt were deposited at different substrate temperatures for each presented sample: we applied room temperature (RT),  450\,$^\circ$C and 600\,$^\circ$C. After the Pt layer deposition a annealing process similar as for the Fe layer was performed at the corresponding growth temperature.

\subsection*{Structural analysis}

\begin{figure}[ht]
\includegraphics[width=0.9\textwidth]{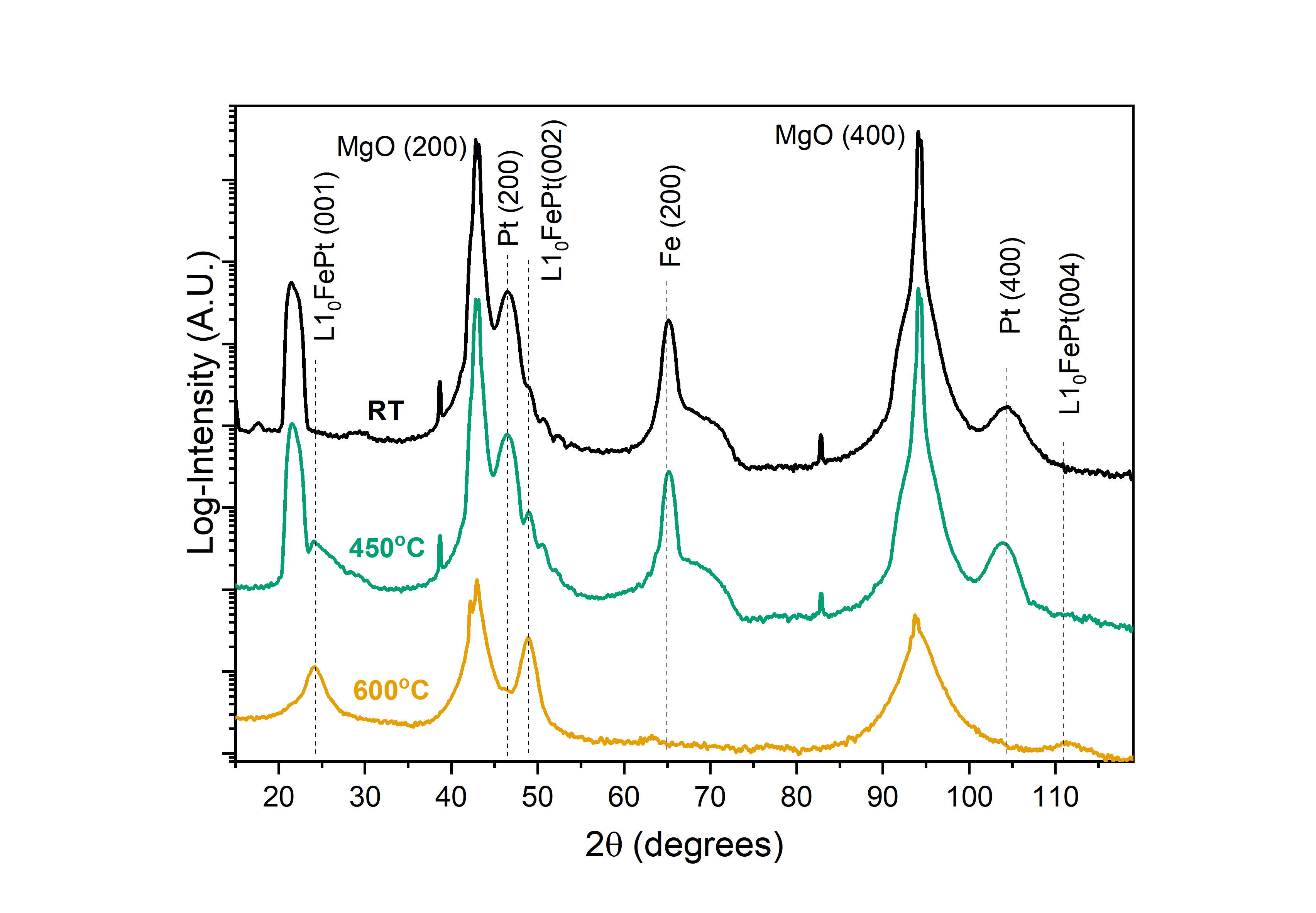}
\caption{\label{fig:xrr} XRD diagrams of the samples with the Pt layer grown at RT, 450$^\circ$C and 600$^\circ$C. The spectra have been vertically shifted for clarity. Epitaxial relationship for the MgO/Fe/Pt is revealed for the RT sample, while the interlayer L1$_{0}$-FePt alloy is induced by an increased growth temperature.}
\end{figure}

Initially, structural investigations were performed with the help of X-ray diffraction (XRD). Figure \ref{fig:xrr} presents XRD patterns of the three Fe/Pt samples. The notation of RT, 450\,$^\circ$C and 600\,$^\circ$C, that will be used throughout the text, refers to the different growth temperature of the Pt layer. For the RT sample, diffraction peaks from the (200) crystal planes of the MgO substrate emerge at 2$\theta$ = 43.01$^{\circ}$  and the related Fe (200) peaks are observed at 65.12$^{\circ}$. Moreover, the Pt (200) peak emerges at 46.31$^{\circ}$. The obtained values for both Fe and Pt are  very close to theoretical peak positions indicative of an almost strain free growth along the growth direction. The preliminary structural characterization confirms that Fe/Pt layers grow epitaxially on MgO\,(100) substrates, following the Bain orientation 
where the \textit{bcc} Fe lattice is rotated by 45$^\circ$ in-plane, with respect to the \textit{fcc} MgO and Pt lattices, to minimize the lattice mismatch~\citep{apl2013,KARFARIDIS2020137716}. The Fe lattice is  rotated in-plane by  45$^\circ$ with respect to both MgO and Pt. The XRD pattern of the 450\,$^\circ$C sample confirms the epitaxial relationship between Fe and Pt layers, however the rise of an additional peak  at 49.1$^{\circ}$ suggests a modified bilayer. Furthermore, the XRD graph for the 600$^\circ$C sample reveals the almost absence of Fe and Pt peaks. The appearance of new diffraction peaks at the 24.07$^{\circ}$, 49.1$^{\circ}$ and 112.12$^{\circ}$ correspond to the (001), (002) and (004) \textit{fct} lattice planes of the ordered L1$_{0}$-FePt, indicating that the alloy phase is now extended in a large part of the volume of the sample.

To further investigate the findings of the XRD analysis, we have implemented nanoscale structural and chemical analysis of the sample grown at 450$^\circ$C, employing  high resolution transmission electron microscopy (HRTEM), Z-contrast high resolution scanning transmission electron microscopy (HRSTEM) and energy dispersive X-ray spectroscopy (EDXS). Cross-sectional TEM specimens were initially wedge polished and finally thinned to electron transparency by Ar$^{+}$ ion-milling in the Gatan PIPS. HRTEM experiments were carried out in a Jeol 2011 UHR TEM, while HRSTEM and STEM-EDXS were performed in a Jeol F200 cold FEG TEM/STEM, both operated at 200 kV.  HRTEM imaging established the formation of an interlayer (IL) at the Fe/Pt interface, as shown in Fig. 2(a), where a characteristic periodic intensity modulation is observed and indicated by arrows. The IL extends on either side of the Fe/Pt interface to a thickness of 5-6 periodicities. The alternating contrast of the IL could be related to the occurrence of a Fe$_{x}$Pt$_{1-x}$ alloy phase that comprises a superlattice structure with high chemical ordering, such as the L1$_{0}$-FePt structure, when projected along the [100] or [010] axes~\citep{Masaaki_2016}. While the interplanar \textit{d}-spacing values of the resolved (020) and (011) crystal planes in MgO/Pt and Fe were found in agreement with their cubic bulk counterparts, suggesting a strain-free configuration, the IL exhibits a tetragonal symmetry. As shown in the high magnification HRTEM image in the left inset of Fig. 2(a), the \textit{d}-spacing values of the IL were 0.193 $\pm$ 0.002 nm along the lateral direction, and 0.186$\pm$0.002\,nm along the growth direction, which are almost identical to the 0.1927\,nm and 0.1856\,nm \textit{d}-spacing bulk values of the (200) and (002) planes of the \textit{fct} L1$_{0}$-FePt lattice, implying an epitaxial growth of the IL phase. 
The average \textit{d}-spacing values were determined by recording intensity profiles along the aforementioned directions, averaged over the whole areas of interest. Moreover, in Z-contrast HRSTEM imaging of the IL (Cs = 0.5\,mm, convergence angle 20\,mrad, detector semi-collection angles 33.4-122\,mrad), the alternating scattering contrast reveals the presence of chemical ordering in monolayer-scale along the growth axis, as illustrated in the top right inset of Fig. 2(a). There, heavier atoms present brighter contrast than lighter ones. In order to associate the intensity modulation with the projection of alternate atomic stacking of Fe and Pt species along the [001] growth axis of the L1$_{0}$-FePt structure, HRSTEM image simulations were performed with the STEMSIM algorithm~\citep{Rosenauer2007}, using the multislice method with absorptive potential approximation. As input, the \textit{fct} L1$_{0}$-FePt structural model was used, projected along the [100] zone axis, with foil thickness up to 20\,nm. Image simulations [bottom right inset of Fig. 2(a)] confirmed that, under our experimental imaging conditions (near Scherzer defocus and $\approx$ 6\,nm foil thickness), the observed intensity modulation is due to the periodic succession of pure Fe and Pt atomic layers of the projected L1$_{0}$-FePt lattice.
Combining all these findings, the 1.9-2.2 nm thick IL is clearly identified as an ordered \textit{fct} L1$_{0}$-FePt alloy, with lattice parameters \textit{a} = 0.386\,nm and \textit{c} = 0.372\,nm  (\textit{c}/\textit{a} = 0.96), which is epitaxially grown following the scheme [100](010)MgO//[100](011)Fe//[001](010)L1$_{0}$-FePt//[100](010)Pt. Further insight, regarding the chemical topography at the Fe/Pt interface, was obtained by STEM-EDXS observations [Fig. 2(b)] showing an interdiffusion of the Fe and Pt layers. Indeed, the intermixing of both elements at the Fe/Pt interface seems to trigger the formation of the L1$_{0}$-FePt IL at 450$^\circ$C and above.

\begin{figure}
\includegraphics[width=0.75\textwidth]{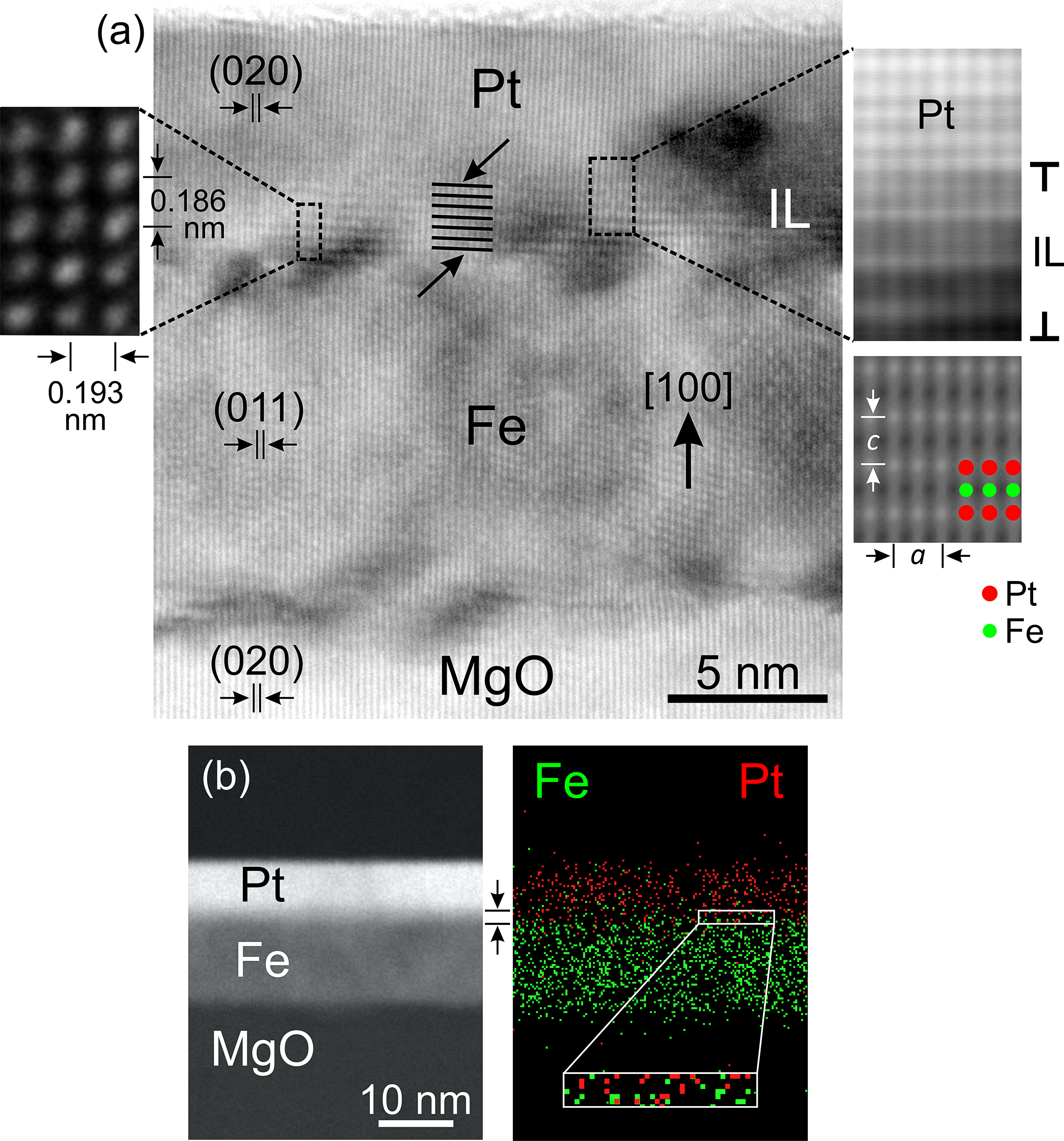}
\caption{\label{fig:tem-GR} (a) Cross-sectional HRTEM image of the 450$^\circ$C sample, along the [001]MgO-Pt/[0-11]Fe zone axis. The out-of-plane crystal planes of MgO, Fe and Pt are indicated, while at the Fe/Pt interface, an IL with periodic intensity is denoted by arrows. The inset on the left hand side illustrates  part of a high magnification HRTEM image obtained from the IL region, showing that it structurally complies with the tetragonal lattice of the L1$_{0}$-FePt alloy with the c-axis along the growth direction. The inset on the right hand side is an  HRSTEM image obtained from the IL region, including the transition to the Pt layer. The alternating  contrast of the IL in the HRSTEM image is attributed to the chemical ordering of the L1$_{0}$-FePt structure, as confirmed by the corresponding HRSTEM image simulation given directly. The [100] projected L1$_{0}$-FePt structural model is superimposed on the simulated image, where the Fe and Pt atoms are indicated, depicting the successive ordering of pure Fe and Pt atomic layers. (b) STEM image, showing the MgO, Fe, IL (arrows) and Pt layers along with the corresponding EDXS map of the distributions of Fe and Pt elements, revealing their intermixing at the Fe/Pt interface, that allows the formation of an L1$_{0}$-FePt alloy IL.}
\end{figure}

\newpage

\subsection*{THz time domain spectroscopy}

\begin{figure} [ht]
   \begin{center}
   \includegraphics[width=0.8\textwidth]{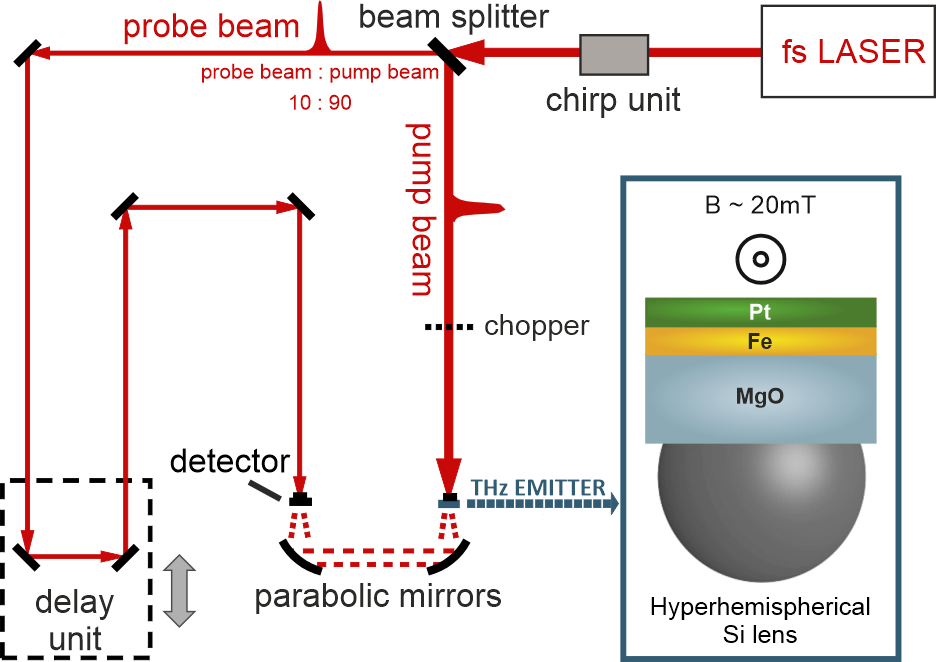}
   \end{center}
   \caption[example] 
   { \label{fig:schematic} 
Schematic of the used Time-Domain THz Spectroscopy (THz-TDS) measurement setup. The spintronic bilayer is used as the source of the THz radiation. As shown in the inset, the fs-laser pulse meets the sample from the Pt side and the THz pulse travels through the metal layers and the substrate and then is collimated by an attached Si-lens. That way, the collection efficiency of our system increases significantly up to 30 times. The exchange of the STEs does not influence the THz detection since the beam path is fixed and the samples stay always on focus allowing our system to compare different samples.}
 \end{figure}

The THz experiments with the Fe/Pt heterostructures were performed with a standard terahertz time domain spectroscope (THz-TDS), where the heterostructures were used as THz emitters. The system is described in detail in Ref. [\citep{Torosyan2018}]. The core of the system is a femtosecond Ti:Sa laser that produces 22\,fs optical pulses at a wavelength of 800\,nm with a repetition rate of 75\,MHz and a typical output power of 500\,mW, as shown in Fig.~\ref{fig:schematic}. 

\begin{figure}[ht]
\includegraphics[width=0.7\textwidth]{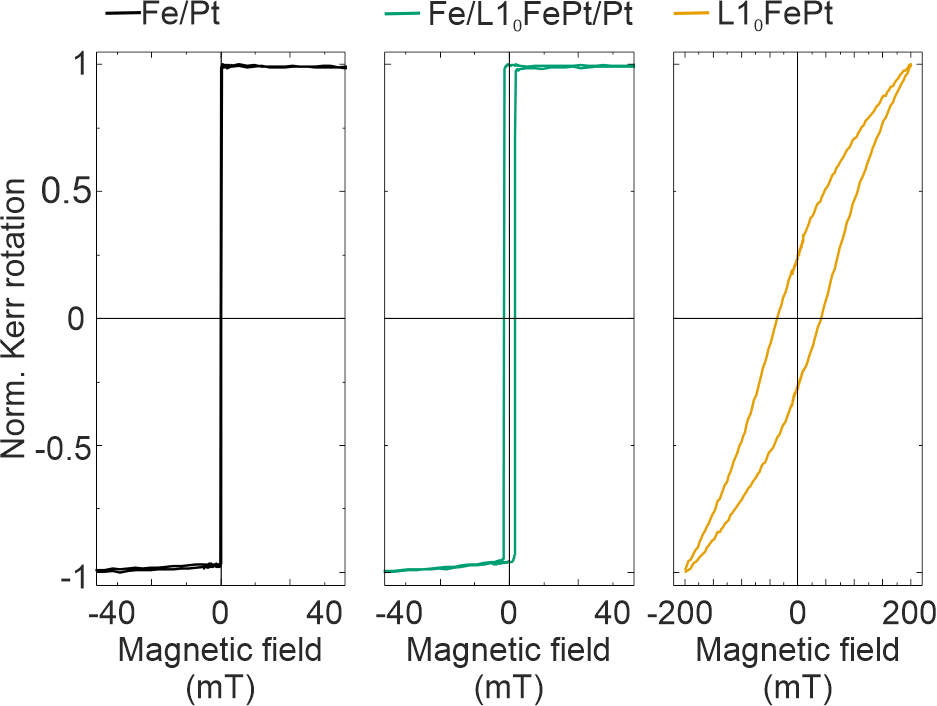}
\caption{\label{fig:moke} Hysteresis curves, recorded with longitudinal Magneto-Optical Kerr Effect for the three samples grown at RT,  450 and 600~$^\circ$C.  The Fe/L1$_{0}$-FePt (2\,nm)/Pt (middle graph) has comparable magnetization reversal with respect to the Fe/Pt bilayer (left graph) with slightly higher coercivity. The external magnetic field is applied along the easy magnetic axis of the samples. When the alloy is the prominent phase, as for the sample grown at  600~$^\circ$C, there is a drastic change in the magnetisation reversal (right graph). There, a magnetic hardening of the sample with large coercivity and almost no saturation up to the maximum available external field (200\,mT) are observed.}%
\end{figure}

The probe beam was used to excite a photoconductive antenna (PCA) with a dipole length of 20\,\text{$\mu$}m acting as THz detector. The spintronic emitter is magnetized by an external magnetic field of maximum available value of 20\,mT, that was able to saturate the two samples grown at RT and at 450$^\circ$C along the magnetic easy axis direction, but not  large enough to saturate the sample grown at 600$^\circ$, see Fig.~\ref{fig:moke}. The direction of the external magnetic field  was perpendicular to the direction of the incident pump beam. The external field determines the polarization plane of the generated THz waves. The Fe/Pt bilayer emits THz pulses into the free space in the shape of a strongly divergent beam. The pump beam was focused onto the emitter from the Pt side and a hyperhemispherical Si-lens was attached to the substrate of the emitter to collimate the beam, see Fig.~\ref{fig:schematic}. The so-formed conical THz beam is led to the PCA detector via THz optics. To guarantee comparable experimental conditions, the alignment of the THz optics and the detector is not changed during the exchange of the spintronic emitters. Since the lateral layer structure of the heterostructures is homogeneous, the substrates and total thicknesses are the same and as the position of the pump beam focus stays constant. Thus, the exchange of emitters does not influence the THz beam path and therefore the measurement sensitivity, rendering the relative measurements comparable.

\begin{figure}
\includegraphics[width=0.68\textwidth]{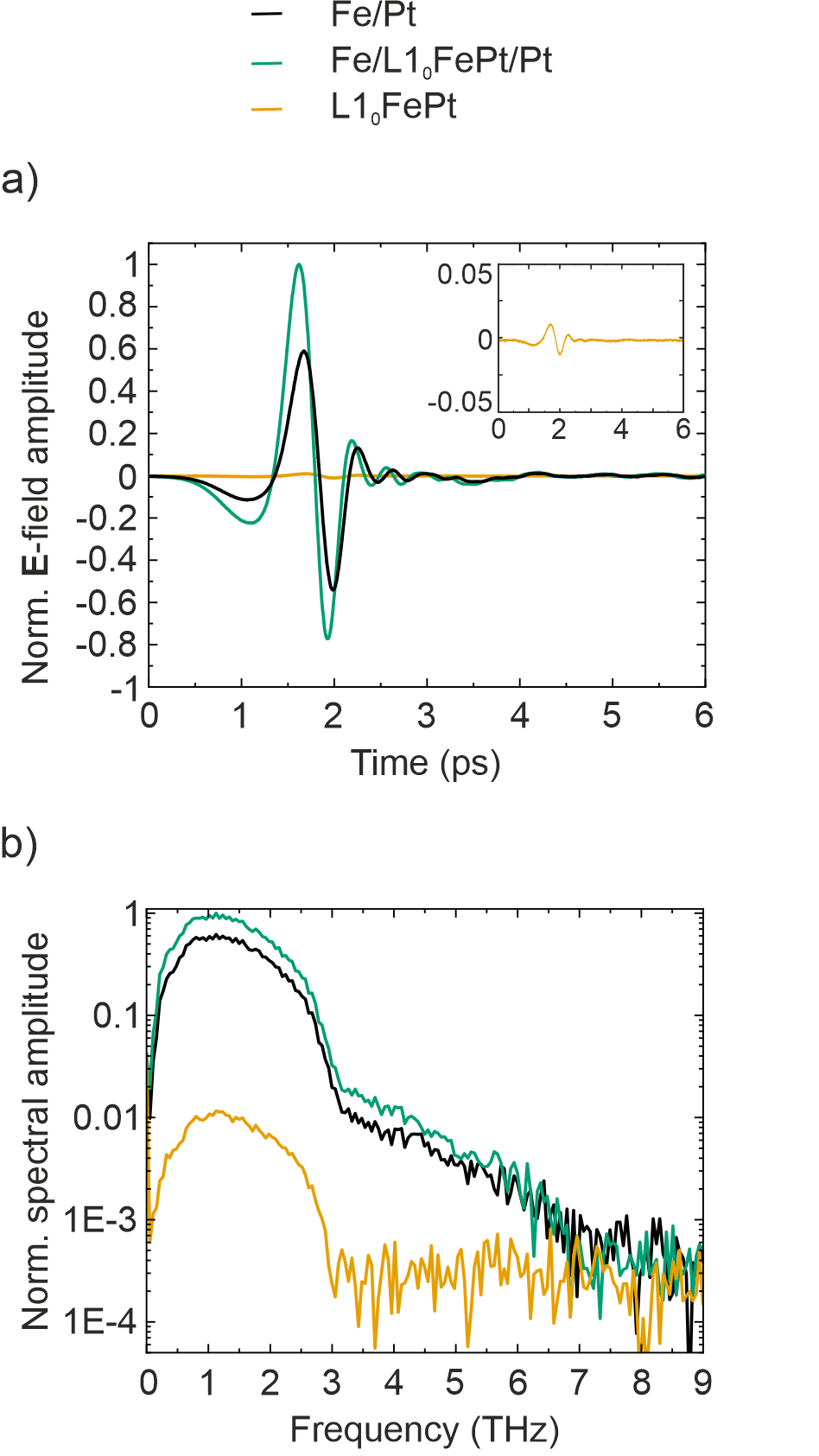}
\caption{\label{fig:pulses} THz emission of the Fe/Pt bilayers with modified interface layers as indicated. The samples are illuminated from the Pt side and the signal is detected from the substrate side with an attached Si-lens. (a) Time-domain THz traces of the pulse train. Interestingly, the sample in the form of Fe/L1$_{0}$-FePt (2\,nm)//Pt exhibits the highest signal. (b) Frequency spectra of the pulse traces as shown in (a) obtained by fast Fourier transformation.}
\vspace{2mm}
\end{figure}

The THz emitted pulses for the 3 different samples grown at RT, 450$^\circ$C and 600$^\circ$C are shown in Fig.~\ref{fig:pulses}(a). The measurements were performed at RT under dry air conditions. The recorded voltage is proportional to the momentary electric field amplitude of the THz wave. The THz spectral amplitude can be obtained by Fourier analysis. The fast Fourier transform (FFT) of the time-trace signal is shown in Figure \ref{fig:pulses}(b). The bandwidth of the PCA detector with a 20 $\mu$m dipole length is limited to a minimum frequency of 100\,GHz and a maximum frequency of 8 THz. While the lowest measurable frequency is only limited by the dipole length of the PCA (longer dipole metallizations allow for the detection of lower frequencies), the detection of higher frequencies is limited by the strong phonon resonances of the GaAs substrate material of the PCA (absorption of the THz radiation).

In Fig.\ref{fig:pulses} (b) above 3 THz the well known strong THz absorption of MgO is visible due to the MgO substrate.  Furthermore, the absorption around 8 THz in the GaAs-detector antenna limits the bandwidth. Besides, there is no need for signal correction due to substrate absorption, since the samples are grown on substrates cut from the same wafer and can be well compared. The maximum frequency measured in these experiments is determined by the frequency response of the dipole antenna of our photo-conductive switch detector and in our case is 1.2\,THz for all of our samples. 

The striking difference between the samples is the amplitude of the observed THz radiation. The largest signal is emitted from the Fe/L1$_{0}$-FePt (2\,nm)/Pt sample, and almost twice as large compared to the Fe/Pt (RT) signal. The pulse for the sample grown at 600$^\circ$C  is clearly present but has a negligible amplitude, see inset Fig.\ref{fig:pulses} (a). Although we are not able to completely saturate the sample grown at 600$^\circ$C (minor loop, see Fig.\ref{fig:moke}), the remanent magnetization value at 20\,mT should in principal be able to provide a detectable THz signal, if the bilayer structure was not strongly modified.  The bandwidth defined at 50\% of the maximum is similar for both the Fe/L1$_{0}$-FePt (2\,nm)/Pt and the Fe/Pt sample. In short, the trilayer Fe/L1$_{0}$-FePt ( 2\,nm)/Pt provides the largest signal while all the samples have similar bandwidth. The epitaxial Fe/Pt grown at RT provides a strong signal which is still smaller than that of the trilayer. Further comparison with a bilayer Fe (12\,nm) / Pt (6\,nm) grown at 300$^\circ$C (see Supplemental Information Fig. S4) reveals once more the superiority of the Fe/L1$_{0}$-FePt (2\,nm)/Pt trilayer.
Although the initially relative high Fe and Pt thicknesses are expected to generate a reduced THz emission with respect to thinner samples, (for example  the optimized Fe (2\,nm)/ Pt (3\,nm) provides 10 times larger amplitude with respect to an optimized Fe (12\,nm) /Pt (6\,nm) \citep{Torosyan2018}), they allow us to minimize the large influence of thickness variations on the THz emission, since thicker layers are insensitive to layer thickness variations of $\approx$ 0.5-1\,nm \citep{Torosyan2018, Seifert2016}. This permits us to observe the influence of the extra thin alloy interlayer.

Last, as the case of the sample grown at 600$^\circ$C shows, when the L1$_{0}$-FePt is much thicker and dominates the whole structure, the THz emission is not enhanced. The low signal for the sample grown at 600$^\circ$C could originate from a THz polarisation signal which is not perpendicular to the magnetization but only to a small projection of the magnetization along the applied magnetic field. However, this is not the case. The sample grown at 600$^\circ$C is magnetically isotropic in-plane when we apply magnetic fields up to 200\,mT as longitudinal MOKE magnetometry reveals (see Supplementary Information Fig. S3). For the maximum applied field, coercivity as well as remanent field remain constant independently of the orientation of the external magnetic field. The isotropic distribution of magnetization shows that the low THz signal of the sample grown at  600$^\circ$C does not depend on the measurement geometry.

\section*{Discussion}

In order to understand the evolution of the THz amplitude, the various factors that can influence the  THz emission need to be properly taken into account.  The far-field THz electric field amplitude $E_{\rm{THz}}$, that is experimentally measured, is proportional to the time-derivative of the local charge current $\mathbf{j}_{\rm{c}}$ that can be written in the time-domain as:
\begin{equation}
\mathbf{E}_{\text{THz}} \propto \frac{\partial \int \mathbf{j}_{\text{c}}dz}{\partial t}
\label{equation 1}
\end{equation}
 where the charge current $\mathbf{j}_{\rm{c}}$ is induced by the inverse spin Hall effect and is proportional to the spin-current  $\mathbf{j}_{\rm{c}}$ $\approx$ $\Theta_{\rm{SHA}}\cdot$ $\mathbf{j}_{\rm{s}}$. 
 Further, the integral over the thickness $dz$ of the non-magnetic layer corresponds to the summation over all the emitting dipoles. $\Theta_{\rm{SHA}}$ is the spin Hall angle. 
 The diffusion of the charge current in the NM layer depends on its spin diffusion length, $\lambda_{\mathrm{sd}}$, and the longitudinal conductivity $\sigma_{\mathrm{long}}$ ~\citep{Torosyan2018, Seifert2016}. 
 Equally important, a critical parameter turns out to be the interface transmission $T$ accounting for the spin current transmission probability at the FM/NM interface, and the characteristic spin-flip times \citep{Papa2019,Jaffres2020}. 
 Therefore, the strength of the THz emission is expected to depend on the product $\propto\,\Theta_{\mathrm{SHA}}\cdot \sigma_{\text{long}}\cdot \lambda_{\text{SD}}\cdot T$.
In the comparison between Fe/Pt and Fe/L1$_{0}$-FePt/Pt,  spin Hall angle $\Theta_{\rm{SHA}}$ and $\lambda_{\rm{SD}}$ are considered to  be kept constant since they refer to  the Pt layer that is mainly unchanged. The alloying of the first Pt atomic layers with Fe at the interface seems to slightly influence  the electrical resistivity  of the samples. Sheet resistivities have been measured with the four-point Van der Paw technique at RT and have given values of: $\rho_{\rm{RT}}= 1.79 \times 10^{-7} \Omega $m, $\rho_{450}= 1.72 \times 10^{-7} \Omega$m, $\rho_{600}= 7.18 \times 10^{-7} \Omega$m. Therefore, the alloying at the interface for the sample grown at 450$^\circ$C seems to even have a reduced resistivity compared to the sample grown at RT, while the strong alloying for the sample grown at 600$^\circ$C modifies drastically its electrical properties. In this context the enhanced THz emission suggests that the main factor for the enhancement compared to the RT-sample is the interface transmission $T$. The large value of $T$ for the L1$_{0}$-FePt interlayer of $\approx$2\,nm thickness indicates that the alloy enhances the spin asymmetry between the Pt 5d band and the Fe 3d band and leads to an increased THz emission. 

The possible enhancement of the interface transparency was further probed by means of the efficiency of spin-to-charge conversion in radio frequency (ferromagnetic resonance) spin-pumping experiments (FMR-SP). Although the excitation of a spin current in FMR-SP experiments is performed on different time- and energy scales, the interface transmission represents the same physical quantity. 

In spin pumping experiments, the efficiency of the interfacial spin transport at a FM/NM interface is characterized by the so-called spin mixing conductance (SMC) $g^{\uparrow\downarrow}$, where it assumes  immediate spin flip in the NM layer. In this case $g^{\uparrow\downarrow}$ is the parameter that needs to be considered for the additional damping by the spin pumping. However, in spin pumping experiments  the so-called effective spin mixing conductance $g^{\uparrow\downarrow}_{\rm eff}$ is typically determined since factors like the finite resistivity and the finite spin diffusion length in the NM layer and interface morphology influence the spin flow in the NM-layer. $g^{\uparrow\downarrow}_{\rm eff}$, is experimentally determined by measuring the increase in the Gilbert damping parameter  $\alpha$  and comparing it with the damping value of the corresponding single magnetic layer without a spin sink layer~\citep{Conca2016,Christoph2020}. The contribution of the spin current dissipation due to spin pumping can be estimated by calculating the damping enhancement and correlating it to the $g^{\uparrow\downarrow}_{\rm eff}$ according to\citep{Keller2018}:

\begin{equation} \label{eq:conductance}
\Delta \mathrm{\alpha}_{\rm sp}= \mathrm{\alpha}_{\rm Fe/Pt}-\mathrm{\alpha}_{\rm Fe-Ref}= \frac{\gamma\hbar}{4\pi M_\mathsf{s} \:d_{\rm FM}}g^{\uparrow\downarrow}_{\rm eff}.
\end{equation}

where $g^{\uparrow\downarrow}_{\rm eff}$  is the effective SMC, which is controlling the magnitude of the generated spin current, $\gamma$ is the gyromagnetic ratio, $M_\mathsf{s}$ the saturation magnetization of each sample, $d_{\rm FM}$ the thickness of the ferromagnetic layer.

The FMR-SP effect was experimentally measured by using a strip-line vector network analyzer (VNA-FMR)\citep{Conca2016}, see also Supplementary Information Fig.~S1. For this, the samples were placed face down and the S$_{12}$ transmission parameter was recorded. The effective Gilbert damping parameter  $\alpha$ was calculated by the dependence of the resonance linewidth on the frequency. Typical measured absorption FMR spectra and the determination of the damping is shown in Supplemental Information Fig. S2. The obtained values are $\alpha_{\rm{RT}}$= 3.9 $\pm$ 0.65 $\times$ 10$^{-3}$ and $\alpha_{450}$= 59.0 $\pm$ 0.31 $\times$ 10$^{-3}$. It was not possible to measure the sample grown at 600$^\circ$C by means of FMR in the available field and frequency range (9 to 20\,GHz). 

Since the spin current leaving the magnetic layer carries away angular momentum from the magnetization precession, it represents an additional loss channel for the magnetic system and consequently causes an increase in the measured Gilbert damping parameter. By comparing the calculated $\mathrm{\alpha}$-values to that of an Fe single layer of the same thickness that was used as a reference sample ($\alpha_{\mathrm{Fe-Ref}}$= 2.15 $\pm$ 0.15 $\times$ 10$^{-3}$~\citep{Conca2016}) we can calculate the additional contribution of the spin current dissipation due to spin pumping using Eq.~\ref{eq:conductance}. Assuming the saturation magnetization of bulk Fe, we calculated the following $g^{\uparrow\downarrow}_{\rm eff}$ values: $g^{\uparrow\downarrow}_{\rm eff-FePt}$ = 2.21 $\times$ 10$^{19}$\,m$^{-2}$ and
$g^{\uparrow\downarrow}_{\rm eff-Fe/FeL1_{0}-FePt/Pt}$ = 80.95 $\times$ 10$^{19}$\,m$^{-2}$. The obtained value for the Fe/Pt bilayer matches well previous studies of our group on similar thicknesses~\citep{RevModPhys.77.1375,Keller2018}.

The reported $\alpha$- and $g^{\uparrow\downarrow}_{\rm eff}$- values need to be taken with caution for the Fe/FeL1$_{0}$-FePt/Pt sample, since Eq.~\ref{eq:conductance} is valid under the assumption that spin pumping is the dominant mechanism affecting this dependence. However, different factors can influence the experimental estimation of the increase in $\mathrm{\alpha}$ and $g^{\uparrow\downarrow}_{\rm eff}$: first of all Fe/FeL1$_{0}$-FePt/Pt is a trilayer composed of two ferromagnetic layers Fe and FeL1$_{0}$ that are magnetically coupled. This can drive the enhancement in damping and can furthermore induce magnetic proximity effects in Pt layer: the so-called magnetic polarization in the NM layer ~\citep{Caminale2016,Keller2018} can  influence the observed damping and SMC estimation. Other effect are able as well as to strongly influence SMC calculation like spin memory loss effect~\citep{Rojas2014} and the two-magnon scattering effect~\citep{Conca2018,Buhrman_2MS_2019}. The aforementioned effects have been proven to influence the obtained values and even to be able to give large 'unphysical' values of $g^{\uparrow\downarrow}_{\rm eff}$ which do not correspond to a real efficient spin transport.
The reported $g^{\uparrow\downarrow}_{\rm eff-Fe/FeL1_{0}-FePt/Pt}$ value  has some limitations due to many factors that can contribute to this enhancement so it can be only considered as an upper limit for $g^{\uparrow\downarrow}_{\rm eff}$. Despite this fact, the large difference between Fe/Pt and Fe/FeL$1_{0}$-FePt/Pt tends to confirm qualitatively the THz spectroscopy data and support the argument that the interface transparency is largely enhanced by the presence of the L1$_{0}$-FePt interlayer.

\section*{Conclusions}

In conclusion, we have grown a series of Fe/Pt bilayers of initially same thickness and we then intentionally reformed the interface by modifying the  growth and annealing temperature of the Pt layer. At 450 $^\circ$C growth temperature  we have achieved the formation of an ordered L1$_{0}$-FePt alloy phase at the interface. Our findings show that the presence of a 2\,nm thick L1$_{0}$-FePt interlayer promotes the interface transmission and amplifies the THz emission. The unique finding of the increased THz emission in Fe/L1$_{0}$-FePt/Pt structure opens new perspectives in the direction of application of spintronic THz emitters and can further stipulate  theoretical and experimental studies.

\section*{Limitations of the study}

The reported $g^{\uparrow\downarrow}_{\rm eff}$- values need to be taken with caution for the Fe/FeL1$_{0}$-FePt/Pt sample, since Eq.\ref{eq:conductance} is valid under the assumption that spin pumping is the dominant mechanism affecting this dependence.

%\section*{Star methods}
%
%\begin{itemize}
%
%\item{Key resource table}
%\item{Resource availability}
%\item{Method details}
%	\begin{itemize}
%	\item{Ferromagnetic resonance spectroscopy and damping calculation}
%	\item{MOKE magnetometry}
%	\end{itemize}
%
%\end{itemize}
%
%\section*{Supplemental information}
%
%Download:
%
%Document Supplemental: Figures S1-S4.

\acknowledgments % equivalent to \section*{ACKNOWLEDGMENTS}       
L.S. acknowledges financial support of the SFB/TRR 173 Spin+X: spin in its collective environment (project B11), funded by the Deutsche Forschungsgemeinschaft (DFG, German Research Foundation), Project No. 290396061/TRR173. E.Th.P. acknowledges the scientific support from Prof. Georg Schmidt (MLU Halle-Wittenberg) and the collaborative research center SFB/TRR 227.

\section*{Method details}

\subsection*{Ferromagnetic resonance spectroscopy and damping calculation}

The dynamic properties of the Fe/Pt sample series were studied by measuring FMR using a stripline and a Vector Network Analyser (VNA-FMR) setup~\citep{Conca2016},see Supplemental Information, Fig.~S1. The dependence of the resonance linewidth $\Delta$H on the frequency $f_\mathrm{fmr}$ is used to determine the Gilbert damping parameter $\mathrm{\alpha}$: 

\begin{equation}
\mu_{0} \Delta H_{\mathrm{FWHM}} = \mu_{0} \Delta H_{0}+ \frac{ 4 \pi \alpha f_\mathrm{fmr}}{\gamma }
\label{eq:Gilbert}
\end{equation} 
\noindent where $\Delta H_{\mathrm{FWHM}}$ is the linewidth of the resonance measured at full width half maximum (FWHM). The inhomogeneous broadening $\Delta H_{0}$ being related to the film quality. $\mu_{0}$ is the vacuum permeability.  

\begin{figure}[]
    \centering
    \includegraphics[width=0.5\columnwidth]{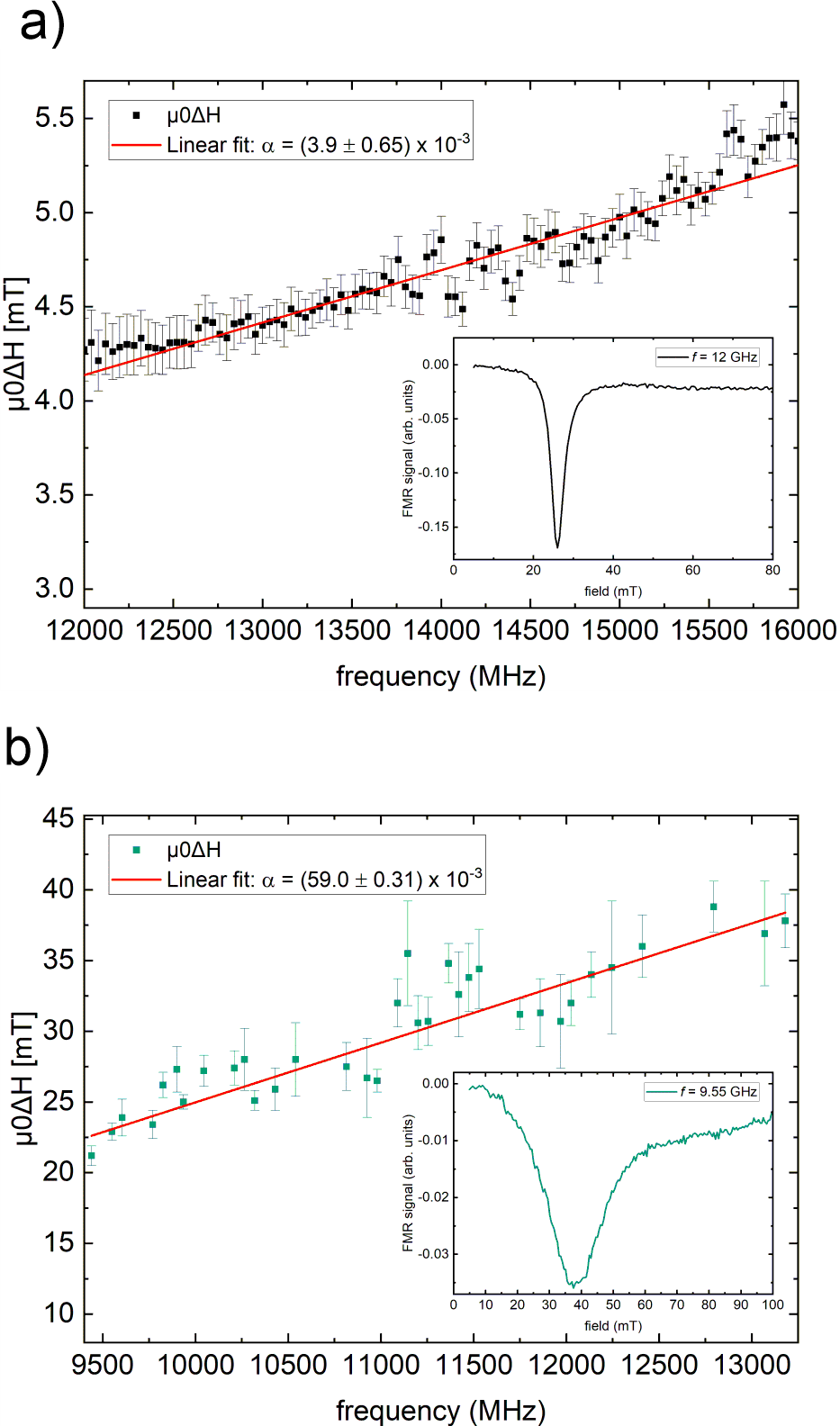}
    \caption{Damping calculation and examples of  FMR absorption spectra for Fe/Pt (Upper Panel) and Fe/L1$_{0}$-FePt/Pt (Lower Panel) samples.  From the dependence of the resonance linewidth on the frequency  the damping parameter $\mathrm{\alpha}$ is determined. The red line is a linear fit according to Eq.~3.  Insets show typical FMR absorption spectra for both samples.}
   \label{fig:FMR-spectra}
\end{figure}

Typical examples of an FMR absortion spectrum (inset) and the experimental determination of $\mathrm{\alpha}$ with the help of Eq.~\ref{eq:Gilbert} for both Fe/Pt and Fe/L1$_{0}$-FePt (2\,nm)/Pt are shown in Fig.~\ref{fig:FMR-spectra}

\subsection*{MOKE magnetometry}

Magnetic hysteresis loops were recorded employing the longitudinal magneto-optical Kerr effect (LMOKE). The excitation was performed with laser light at $\lambda$ = 660\,nm and at 45$^{\circ}$ oblique incidence, yielding information about the in-plane magnetization of the films. Furthermore, by rotating the
magnetic field in-plane from 0$^{\circ}$ to 360$^{\circ}$ the in-plane magnetic anisotropy of the samples was probed. In Supplemental Information Fig.~S3 the in-plane measurements for the alloyed sample grown at 600$^{\circ}$C are presented. No change in the magnetization reversal was observed. The maximum available external magnetic field was not able to fully saturate the sample.

%
%\section*{Declaration of interests}
%
%The authors declare no competing interests
%
%\subsection*{Data and code availability}
%
%There is no dataset or code associated with this work.

%\bibliography{report}

\end{document}